\documentclass{aa}  
\usepackage{graphicx}
\usepackage{txfonts}
\begin{document}

   \title{Connecting substellar and stellar formation. The role of the host star's metallicity.}

   \subtitle{}

   \author{J. Maldonado
          \inst{1}
          \and
          E. Villaver\inst{2}
	  \and
	  C. Eiroa \inst{2}
	  \and
	  G. Micela \inst{1} 
          }

   \institute{INAF - Osservatorio Astronomico di Palermo,
              Piazza del Parlamento 1, 90134 Palermo, Italy\\
              \email{jesus.maldonado@inaf.it}
         \and
              Universidad Aut\'onoma de Madrid, Dpto. F\'isica Te\'orica,
	                    Facultad de Ciencias, Campus de Cantoblanco, 28049 Madrid, Spain
             }

   \date{Received September 15, 1996; accepted March 16, 1997}

 
  \abstract
   {
   Most of our current understanding of the planet formation mechanism is based 
   on the planet metallicity correlation derived  mostly from solar-type stars harbouring gas-giant planets.
   }
   {
   To achieve a far more reaching grasp on the substellar formation process
   we aim to analyse in terms of their metallicity a diverse  sample of stars
   (in terms of mass and spectral type) covering the whole range of possible
   outcomes of the planet formation process (from planetesimals to brown dwarfs and low-mass binaries).
   }
   {
   Our methodology is based on the use of high-precision stellar parameters derived by our own group
   in previous works from high-resolution spectra by using the iron ionisation and equilibrium conditions.
   All values are derived in an homogeneous way, except for the M dwarfs where a methodology based
    on the use of pseudo equivalent widths of spectral features was used. 
   } 
   {
   Our results show
   that as the mass of the substellar companion increases the metallicity of the host star tendency is to lower values. 
   The same trend is maintained when analysing stars with low-mass stellar companions and a 
   tendency towards a wide range of host star's metallicity is found  for systems with low mass planets.   
   We also confirm that more massive planets tend to orbit around more massive stars.
   }
   {
   The core-accretion formation mechanism for planet formation achieves its maximum efficiency 
   for planets with masses in the range 0.2 and 2 M$_{\rm Jup}$. Substellar objects with higher masses
   have higher probabilities of being formed as stars. Low-mass planets and planetesimals might be formed
   by core-accretion even around low-metallicity stars.
   }
   
   
   \maketitle
\section{Introduction}\label{introduction} 

 Exoplanetary science has succeeded in discovering an astonishing
 diversity of planetary systems. 
 The role of the host star's metallicity in planet formation
 has been largely discussed with the finding that 
 the frequency of giant planets is a strong function of the stellar metallicity
 \citep{1997MNRAS.285..403G,2004A&A...415.1153S,2005ApJ...622.1102F}.
 The planet-metallicity correlation is
 usually interpreted in the framework of the core-accretion model
 \citep[e.g.][]{1996Icar..124...62P}
 as the final mass of cores
 via oligarchic growth increases with the solid density in proto-planetary discs 
 \citep{2002ApJ...581..666K}. 

 Initially found for gas-giant planets around solar-type stars
 \citep{1997MNRAS.285..403G,2005ApJ...622.1102F},
  many works have tried to 
 probe whether the gas-giant planet metallicity correlation also holds
 for other kind of stars as well as other types of substellar objects, e.g. low-mass planets
 \citep{2010ApJ...720.1290G,2011arXiv1109.2497M,2011A&A...533A.141S,2012Natur.486..375B,2015ApJ...808..187B},
 brown dwarfs \citep{2011A&A...525A..95S,2014MNRAS.439.2781M,2014A&A...566A..83M,2017A&A...602A..38M}, stars with debris discs
 \citep{2005ApJ...622.1160B,2006A&A...452..921C,2006MNRAS.366..283G,2009ApJ...705.1226B,2009ApJ...700L..73K,2012A&A...541A..40M,2015A&A...579A..20M,2016ApJ...826..171G},
 evolved (subgiant and red giant) stars
 \citep{
 2005PASJ...57..127S,2005ApJ...632L.131S,2007A&A...475.1003H,2007A&A...473..979P,2008PASJ...60..781T,2010ApJ...725..721G,
 2013A&A...554A..84M,2013A&A...557A..70M,2015A&A...574A..50J,2015A&A...574A.116R,2016A&A...588A..98M},
 low-mass (M dwarf) stars \citep[e.g.][]{2013A&A...551A..36N}, or
 if there are differences in the host star metallicity
 when close-in and more distant planets are present in the system 
  \citep{2004MNRAS.354.1194S,2017arXiv171201035M,2018AJ....155...68W}.
  It should be noted that most of these references refer to radial velocity planets.


 As clear from the references above, previous works focus on particular types of stars and planets and,
 to the very best of our knowledge, a global view of the planet-metallicity correlation
 and its implications on the planet formation process are still missing. This is precisely the goal of this work in which
 we analyse in the most homogeneous possible way a large sample of stars harbouring
 the full range of possible outcomes of the planet formation process (from debris discs to
 massive brown dwarfs) and without any restriction of the host star's spectral
 type (from M dwarfs to early-F) or evolutionary
 status (from main-sequence to giants). The analysis is completed with literature
 data of low-mass binary stars in order to set the results into a general context.

 The paper is organised as follows. Section~\ref{stellar_sample} describes the
 stellar sample.
  The completeness of the planet host subsample is analysed in Sect.~\ref{completness}.
 The analysis of the host star metallicities as a function of the substellar companion
 mass and the mass of the host star is performed in 
 Sect.~\ref{analysis}. 
 The results are discussed in the context of current planet
 formation models in Sect.~\ref{pm_context}.
  A comparison with the results from the {\sc Kepler} mission is performed in Sect.~\ref{kepler_comp}.
 Our conclusions follow in Sect.~\ref{conclusions}.

\section{Stellar sample}\label{stellar_sample}

 Our stellar sample 
 is selected from
 our previous works \citep{2012A&A...541A..40M,2013A&A...554A..84M,2015A&A...577A.132M,2015A&A...579A..20M,2017arXiv171201035M,2016A&A...588A..98M,2017A&A...602A..38M} which might be consulted for further details.
 Briefly, high-resolution \'echelle spectra of the stars were obtained
 in 2-3 metre class telescopes or obtained from public archives.
 Basic stellar parameters
 (T$_{\rm eff}$, $\log g$, microturbulent
 velocity $\xi_{\rm t}$, and [Fe/H])
 were determined by using the code {\sc TGVIT}\footnote{http://optik2.mtk.nao.ac.jp/\textasciitilde{}takeda/tgv/}
 \citep{2005PASJ...57...27T}
 which implements the iron ionisation, match of the curve of growth
 and iron equilibrium conditions.
 Stellar age, mass, and radius were computed from
 {\sc Hipparcos} V magnitudes \citep{1997ESASP1200.....E} and
  parallaxes \citep{2007A&A...474..653V}, when available, using the code
  PARAM\footnote{http://stev.oapd.inaf.it/cgi-bin/param} \citep{2006A&A...458..609D},
	 together with
	   the PARSEC set of isochrones \citep{2012MNRAS.427..127B}.
 For some planet hosts where the PARAM code failed to give a reasonable value we took the mass values from the NASA exoplanet archive
  (specifically from the summary of stellar information table, for all stars but KOI 415 for which we took the value from the KOI stellar properites table).
 For the M dwarfs a different methodology was used. The analysis is based on the use
 of ratios of pseudo equivalent widths of spectral features which are sensitive
 to the effective temperature and the stellar metallicity \citep{2015A&A...577A.132M}.
 Stellar masses and luminosities for M dwarfs were obtained from the derived temperatures
 and metallicities by using empirical calibrations.
 
 The total number of stars in our sample amounts to 551.
 It is composed of 71 F-type stars, 261 G-type stars, 166 K-type stars,
 and 53 early-Ms. Regarding their evolutionary state, 373 are in the main-sequence,
 63 are classified as subgiants, and 115 are giants.

\begin{figure}[htb]
\centering
\begin{minipage}{\linewidth}
\includegraphics[scale=0.42]{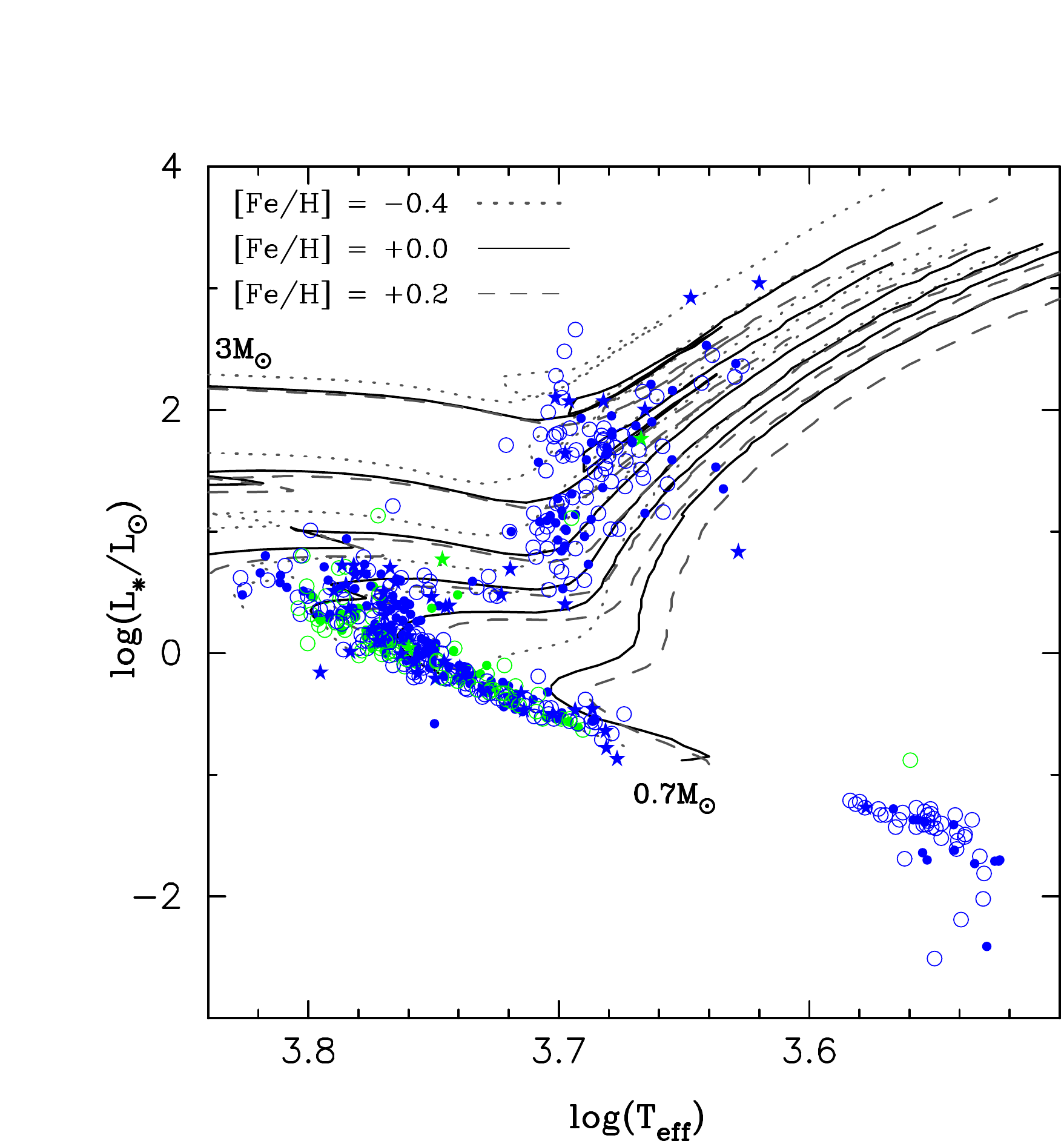}
\end{minipage}
\caption{Luminosity versus T$_{\rm eff}$ diagram for the stars analysed. 
Stars with discs are plotted as green whilst stars without discs are shown in blue.
Stars with planets are shown as filled circles and stars with companions in the brown dwarf
regime are shown in filled stars. Some evolutionary tracks ranging from 0.7 to 3.0 solar masses from 
\cite{2000A&AS..141..371G}
are overplotted. For each mass, three tracks are plotted, corresponding to Z = 0.008 ([Fe/H] = -0.4 dex, dotted lines), Z = 0.019 ([Fe/H] = +0.0 dex, solid lines), and Z = 0.030 ([Fe/H] = +0.20 dex, dashed lines).}
\label{diagrama_hr}
\end{figure}

 The different architectures of the planetary systems harboured by our sample 
 (type and number of substellar objects) are shown in Table~\ref{tabla_sample},  
 while figure~\ref{diagrama_hr} shows the Hertzsprung-Russell (HR) diagram of the stars analysed. 
 Our sample is mainly composed of ``mature'' planetary systems as 
 only  $\sim$ 7\% of our stars have estimated ages younger than 500 Myr.


\begin{table}
\centering
\caption{
  Architecture of the planetary systems in our stellar sample.
}
\label{tabla_sample}
\begin{scriptsize}
\begin{tabular}{lcl}
\hline\noalign{\smallskip}
 Type         & Number & Notes \\
\hline 
 Substellar objects &  345 &  95 multiple systems \\
 (total)            &            &                           \\ 
\hline
 Brown dwarfs				  & 59   & 3 systems with 2 BDs \\
 ( 10 M$_{\rm Jup}$ < m$_{\rm C}\sin i$ < 70 M$_{\rm Jup}$) &  & 5 systems BD + planet \\
\hline
 Low-mass planets                       &     78 &  34 hot (a $<$ 0.1 au) \\ 
 (m$_{\rm C}\sin i$ < 30 M$_{\oplus}$)  &        &  44 cool (a $>$ 0.1 au) \\
\hline
 Gas-giant planets                      &    208 & 34 hot (a $<$ 0.1 au) \\ 
 (m$_{\rm C}\sin i$ > 30 M$_{\oplus}$)  &        &  174 cool (a $>$ 0.1 au) \\
\hline
 Debris disc                            &     99 & 32 Debris disc + substellar object \\
 \hline
\end{tabular}
\end{scriptsize}
\end{table}

\section{Sample completeness}\label{completness}

A deep analysis of the completeness or detectability of our planet host sample is difficult to
overcome as our targets are not selected from any particular exoplanet survey.
Indeed, they were initially selected to address different individual aspects of the planet
formation process (presence of discs, planet formation around evolved stars,
brown-dwarf vs. planet formation, hot vs. cool planets ...).
Our planet host sample includes targets from different radial velocity surveys, e.g.,
the HARPS search for southern extra-solar planets \citep{2004A&A...423..385P};
the Anglo-Australian planet search \citep{2001ApJ...551..507T};
the N2K survey \citep{2005ApJ...620..481F};
the UCO/Lick survey \citep{2006A&A...454..943H};
the the PennState-Toru{\'n} Centre for Astronomy Planet Search \citep{2008ASPC..398...71N};
the retired A stars project \citep{2007ApJ...665..785J}; 
or the list of stars with brown dwarf companions by \cite{2014MNRAS.439.2781M} and \cite{2016A&A...588A.144W};
among others. 
In other words, our planet host sample
comes from a wide variety of planet search programmes with (most likely) different selection criteria,
sensibilities, and biases, sampling significantly different regions of the HR diagram. 

On the other hand, the comparison sample is mainly drawn from the {\sc Hipparcos} 
catalogue  \citep{1997ESASP1200.....E} and was chosen to cover similar stellar parameters as the stars with
detected planets. 
For the sake of completeness, we give in Table~\ref{parameters_table_full} the basic properties of the full sample of stars covered here.
Further details can be found in our previous works (see references above).

 In order to estimate the detectability limits of our planet hosts we proceeded as follows.
 For each star we searched for its corresponding radial velocity curve. Whenever possible, radial
 velocity data was taken from the {\sc NASA} Exoplanet Archive
 \footnote{https://exoplanetarchive.ipac.caltech.edu}. Otherwise, we searched for the data in 
 the corresponding discovery's paper. We were able to recover the radial velocity series for
 89.6\% of our stars with planets. 
 We subtracted the contribution of the known planets to each radial velocity data set by fitting
  a keplerian orbit using the code {\it rvlin}\footnote{http://exoplanets.org/code/} \citep{2009ApJS..182..205W}. The fits were done by fixing the planetary period to the published
   values. When several planets were present around the same star, we subtract them in a sequential way.
    We took into account the fact that data obtained with different instruments might be available for the same
     star.  Once the contribution for the known planets have been removed from the radial velocity datasets,
      we considered the rms of the residuals to be representative of our measurement uncertainty \citep[e.g.][]{2001A&A...374..675E}.

 For each planet host we computed the expected radial velocity semiamplitude due to the presence
 of different types of planets considering circular orbits, as usually done in the literature
 \citep[see e.g.][and references therein]{2011arXiv1109.2497M}.
 We also note that it has been shown that even eccentricities as high as 0.5 do not have a strong
 influence in the planet detection's limits \citep{2002A&A...392..671E,2010MNRAS.401.1029C}. 
 We sampled the planetary mass space
 in logarithmic space, with values ranging from 0.005 to 80 M$_{\rm Jup}$. 
 Regarding the orbital periods, we sampled the orbital frequencies (also in logarithmic space)
 from periods from one to 10$^{\rm 4}$ days.
 For each planet we computed the expected radial velocities keeping the same time as the original observations.
 A total of eleven realizations of the radial velocity, each simulation corresponding to a different phase offset
 (from 0 to 2$\pi$), were performed.
We considered a planet to be detectable around a given star if the rms of the planet's expected radial velocity is larger
 than the rms of the stellar radial velocity residuals in each of the simulated phases \citep{2005A&A...443..337G,2012A&A...542A..18L,2012A&A...545A..87M}.
 We are aware that this is a ``conservative'' approach, i.e, it might overestimate the detection limits for some periods.
 It is however, a fast and robust method, 
 ideal for achieving a quick look an for obtaining an efficient determination of the detection limits
 \citep{2012A&A...545A..87M}.
 It should be noted that for the scope of this work a conservative approach should be preferred in order to obtain robust conclusions.

\begin{figure}[htb]
\centering
\begin{minipage}{\linewidth}
\includegraphics[scale=0.50]{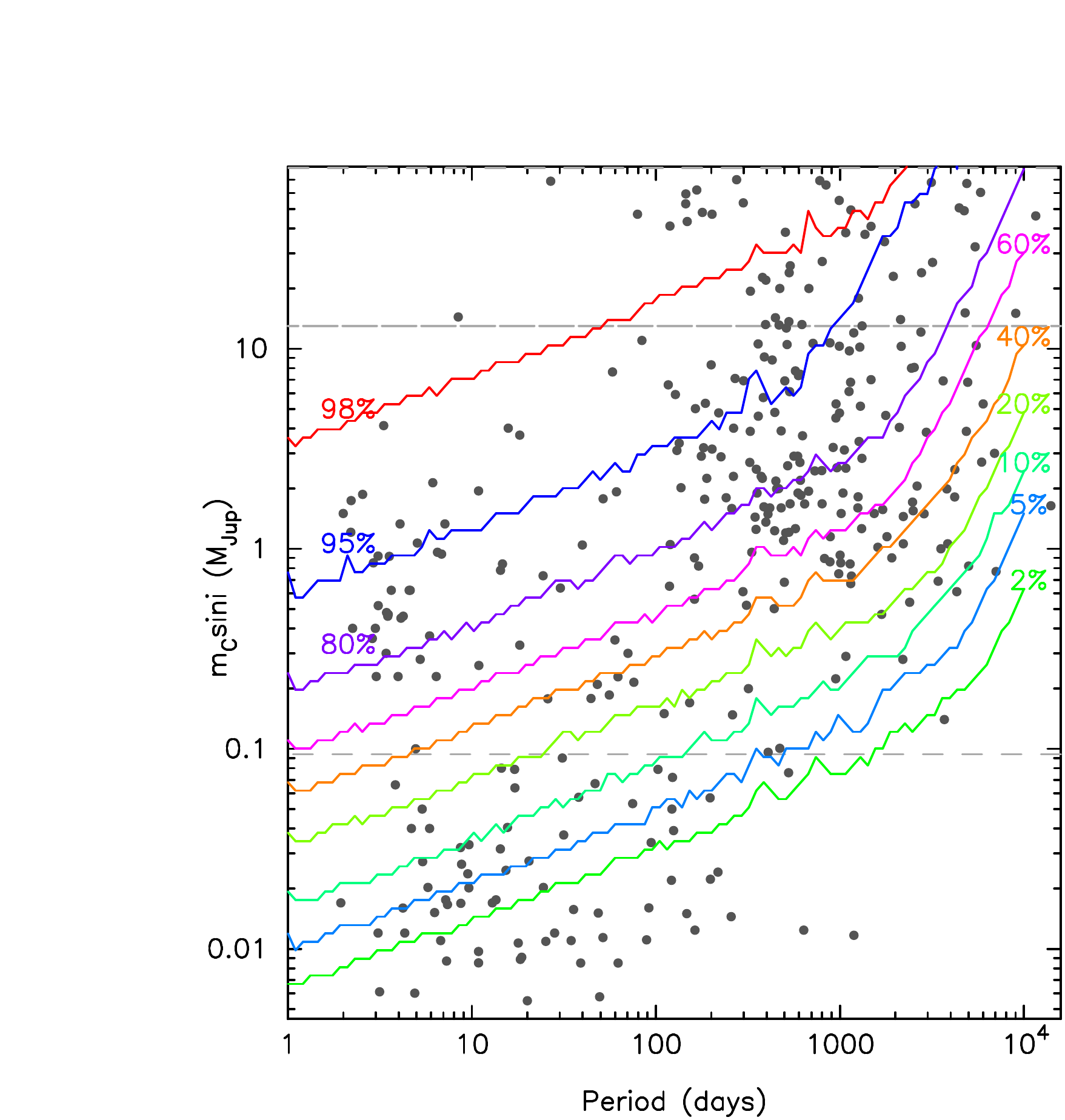}
\end{minipage} 
\caption{ 
Minimum mass versus planetary period diagram. Substellar companions analysed in this work
are shown in grey circles. Detection probability curves are superimposed with different colours.
Horizontal dashed lines indicate
the standard mass loci of 
low-mass planets, and gas-giant planets companions. 
}
\label{detectability}
\end{figure}

 Figure~\ref{detectability} shows the derived detection probability curves. They show for each period, the percentage of stars from our sample
 for which planets with the corresponding minimum mass might be detected, i.e., planets located in the region above the $p$\% curve can be detected in
 $p$\% of our stars.

 Several main conclusions can be drawn from this plot: 
 i) First, most of the long-period planets ($P$ $>$ 100 days) are detectable in approximately  more than 60-80\% of our targets.
 Only for planets with periods longer than $\sim$ 2000 days with a mass of the order of few Jupiter masses and lower, our detectability
 fraction decreases to $\sim$ 40\% and below; 
 ii) Low-mass planets, on the other hand, are detectable only in a small fraction of stars, between 2 and 20\%;
 iii) Finally, it can be seen that planets with the mass of Jupiter and short periods are detectable in practically all stars. 
 
 As noted before our approach is quite conservative, so it is not surprising that some planets are actually located
 in the region under the 2\% probability curve. 
 We will discuss at length the implications of these findings in our analysis in the next sections.

\begin{figure*}[!htb]
\centering
\begin{minipage}{0.33\linewidth}
\includegraphics[scale=0.35]{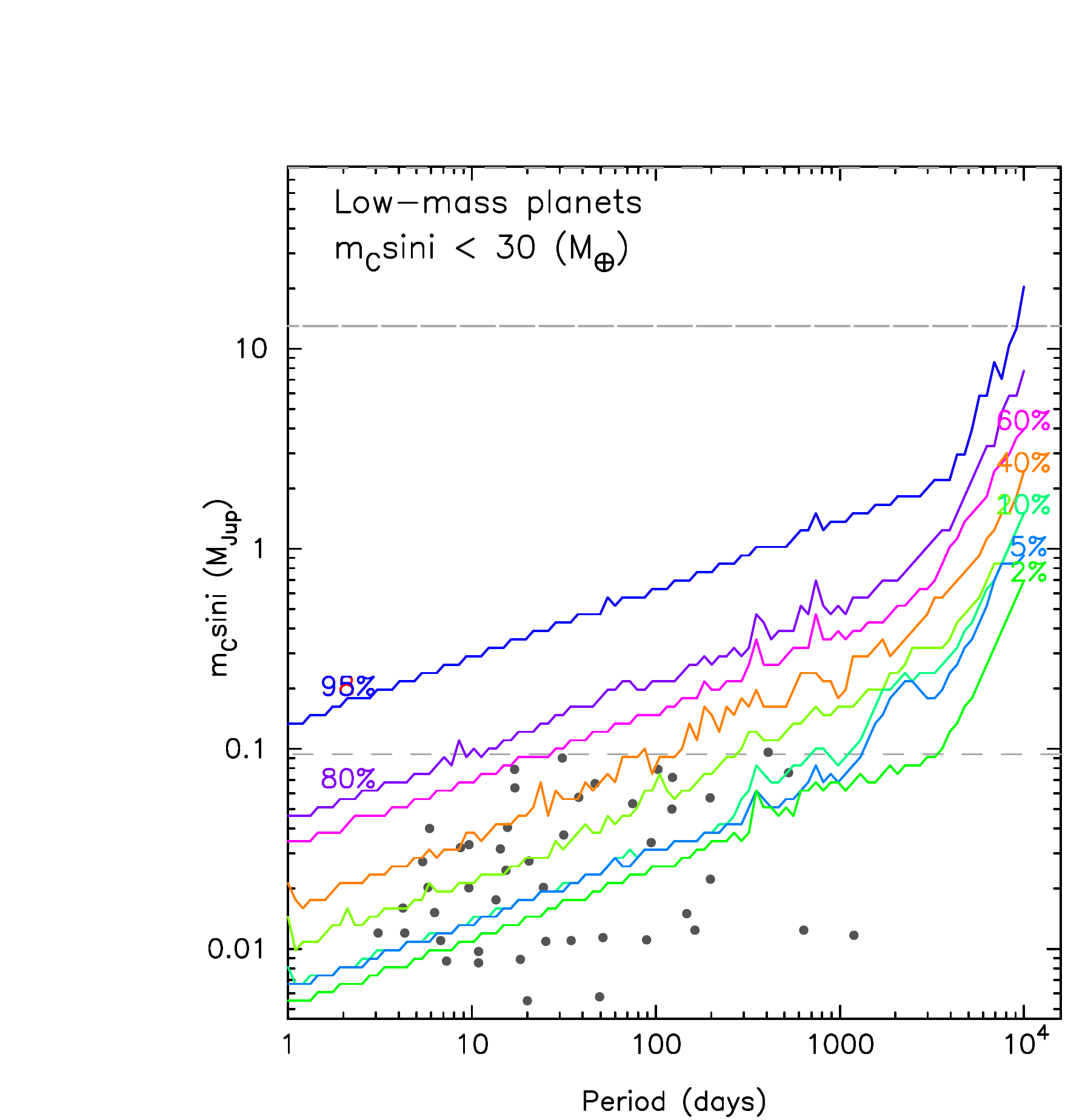}
\end{minipage}
\begin{minipage}{0.33\linewidth}
\includegraphics[scale=0.35]{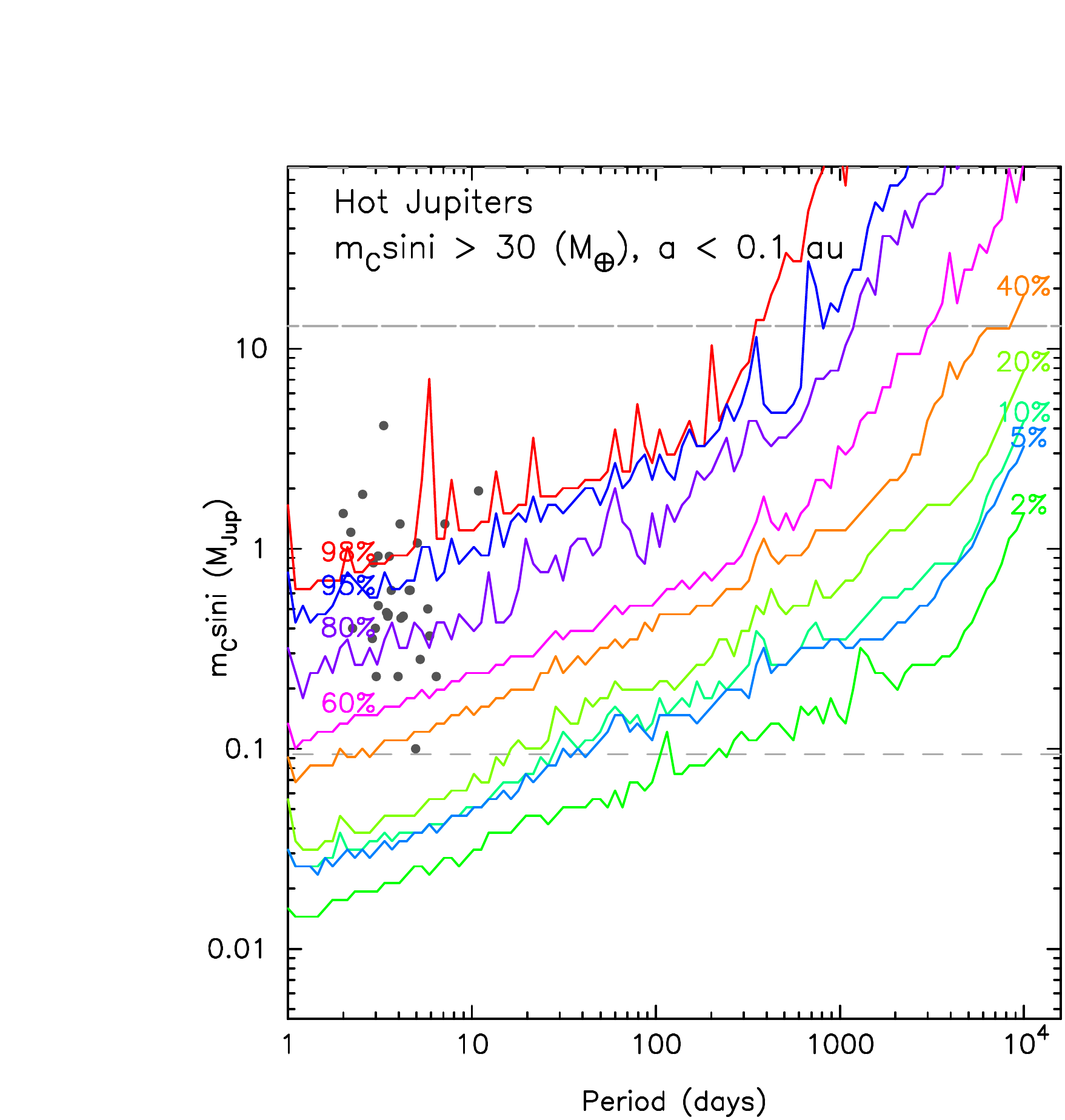}
\end{minipage}
\begin{minipage}{0.33\linewidth}
\includegraphics[scale=0.35]{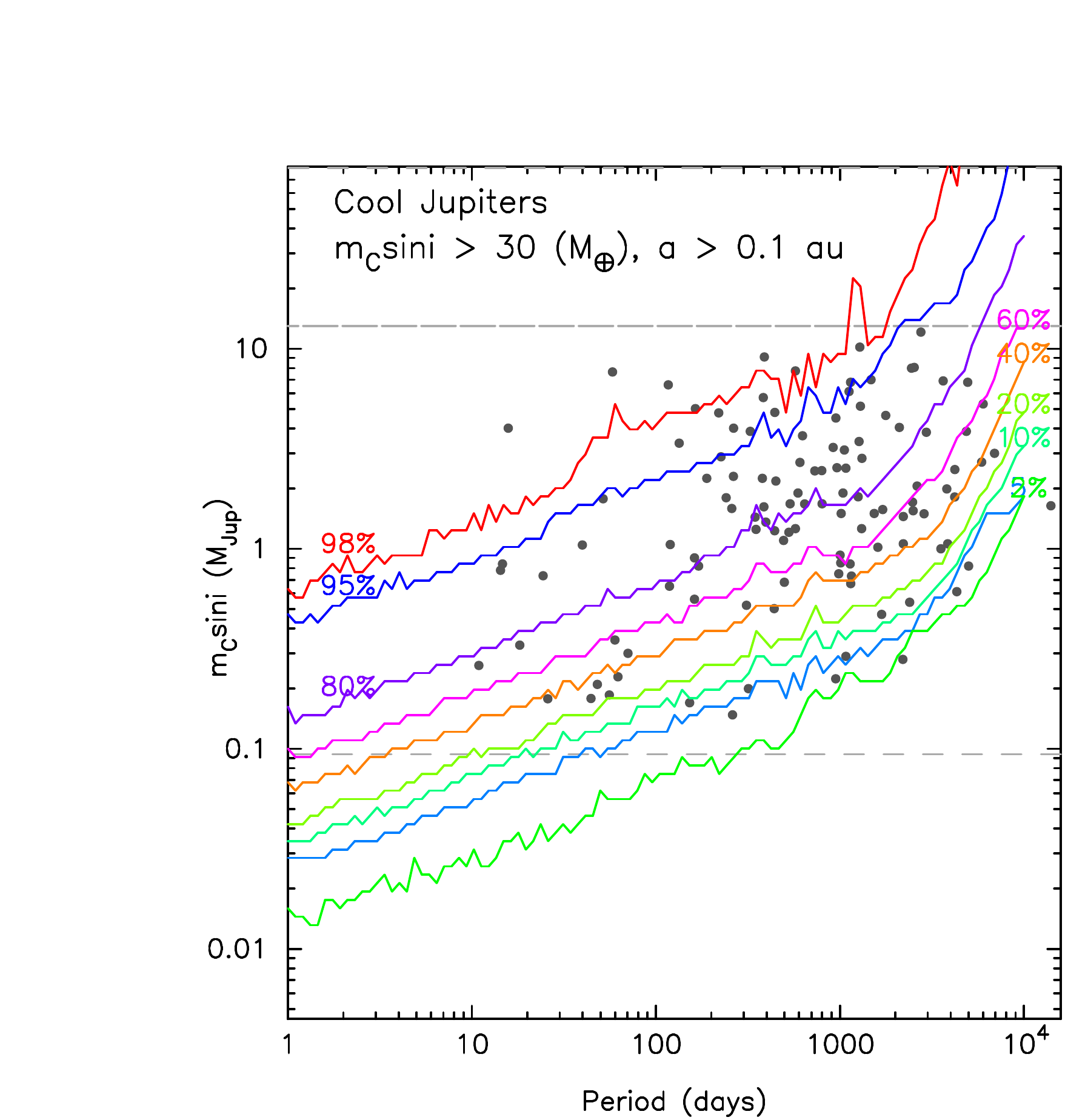}
\end{minipage}
\begin{minipage}{0.33\linewidth}
\includegraphics[scale=0.35]{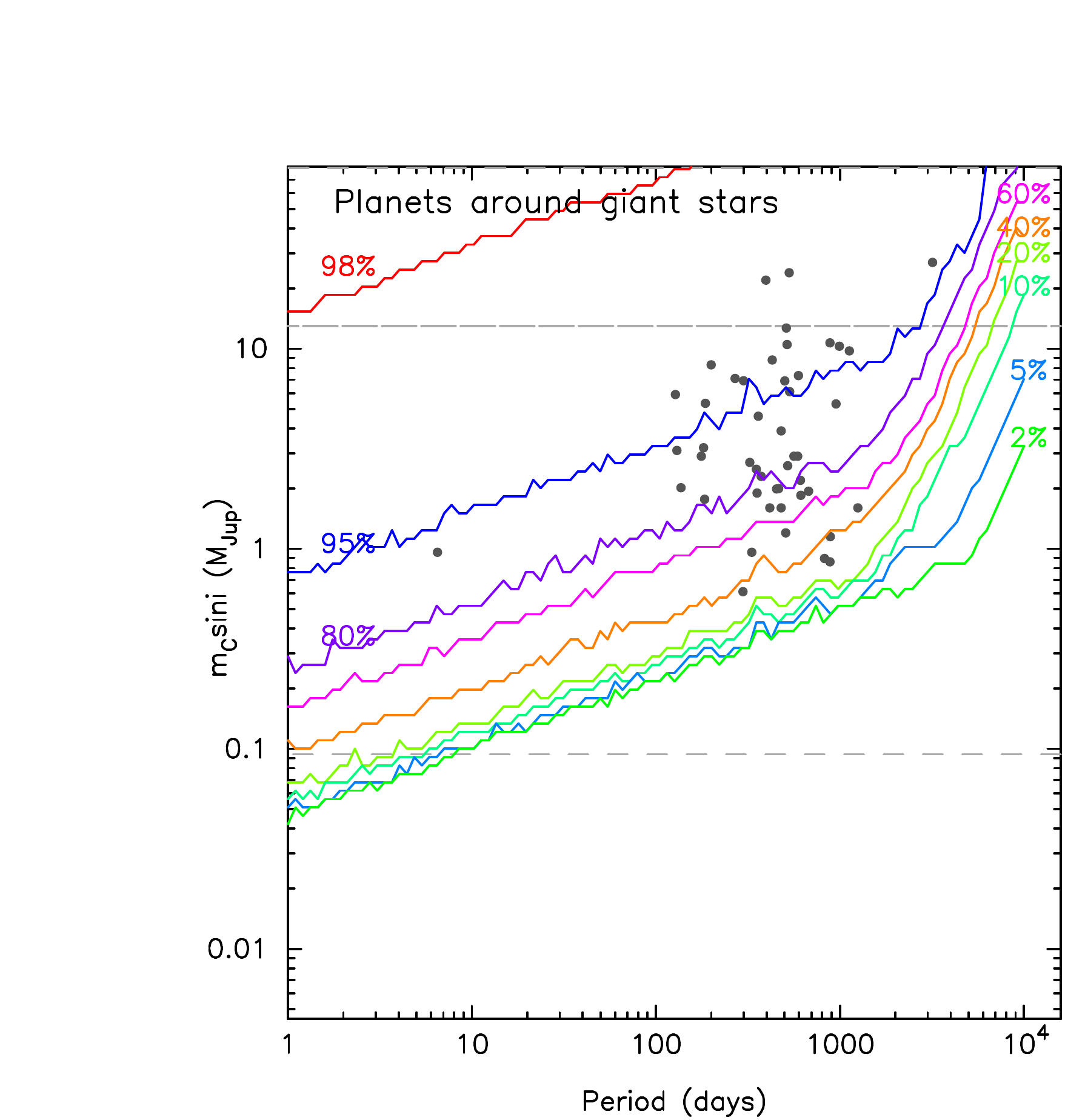}
\end{minipage}
\begin{minipage}{0.33\linewidth}
\includegraphics[scale=0.35]{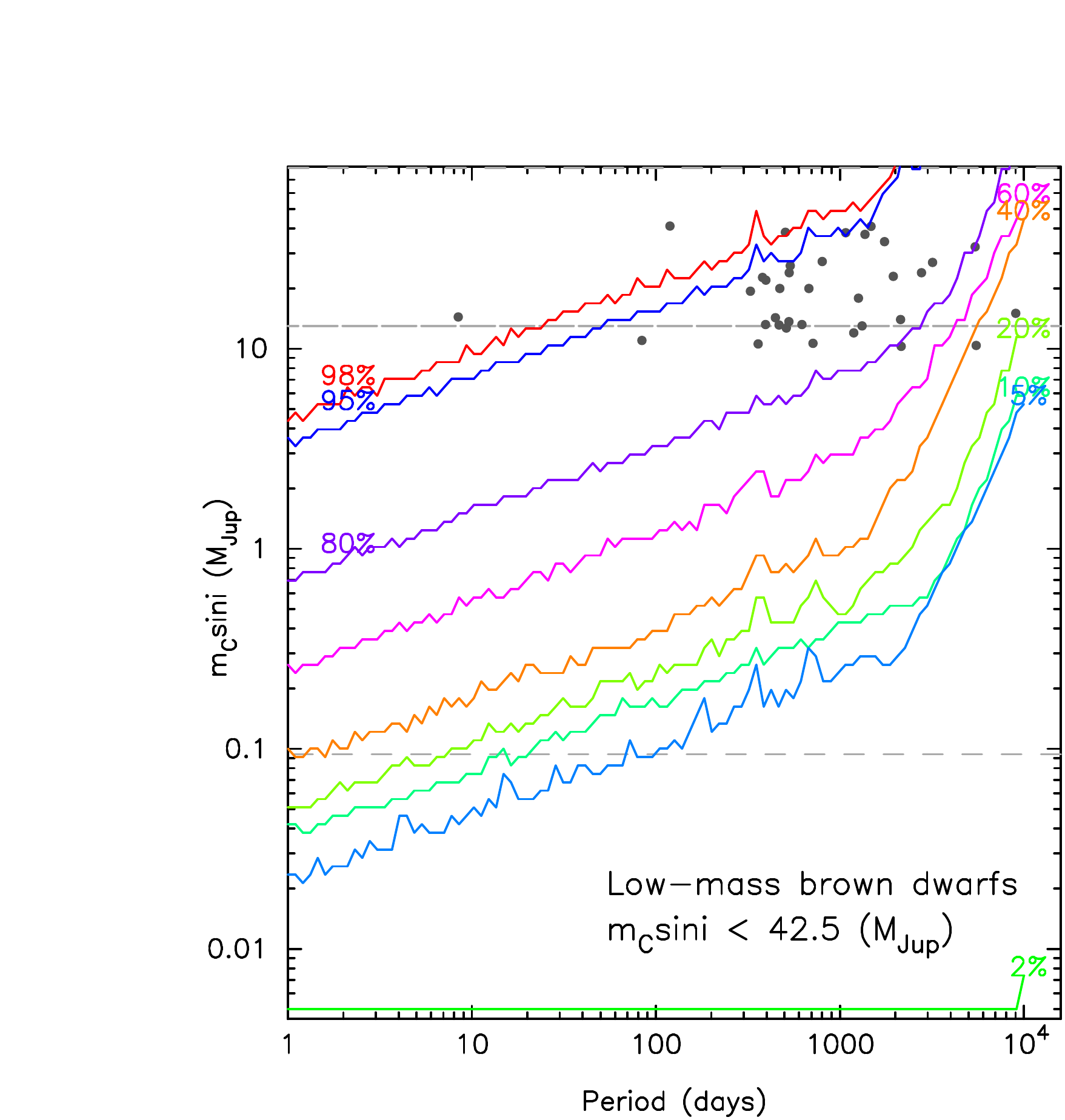}
\end{minipage}
\begin{minipage}{0.33\linewidth}
\includegraphics[scale=0.35]{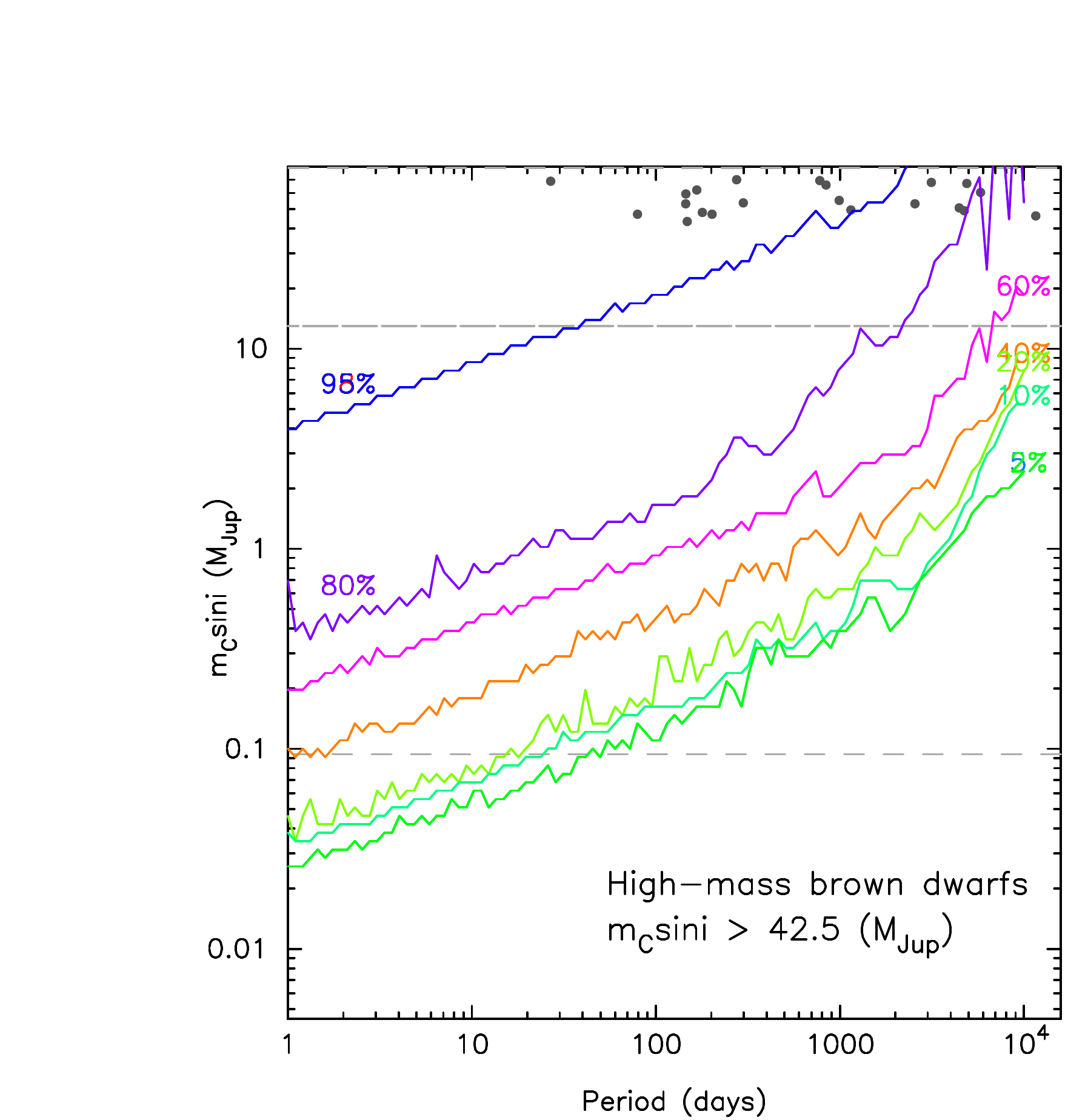}
\end{minipage}
\caption{  
Same as Fig.~\ref{detectability} but for the different subsamples described in 
Sect.~\ref{analysis}.
}
\label{activity_indices}
\end{figure*}

  Figure~\ref{activity_indices} shows the detectability limits for different subsamples of interest,
 see Sect.~\ref{analysis}.
 In addition to the conclusions from the previous figure
 it can be seen the detectability curves for surveys aiming to detect low-mass planets are clearly shifted towards
 lower planetary mass companions, as expected. On the other hand surveys of stars with brown dwarfs and
 specially surveys of giant stars show significant  higher detection limits. 
 Stars with cool and hot Jupiters show nearly identical detection curves. 
%
 Finally, we can see that
 we are mostly insensitive to the presence of small planets (specially evident in the case of surveys around evolved stars) 
 which demonstrates the need of dedicated 
 intensive long-term surveys (and probably the development of specific techniques to deal with the stellar noise problem) 
 in order to detect this kind of planets.


\section{Analysis}\label{analysis}

\begin{figure}[htb]
\centering
\includegraphics[scale=0.42]{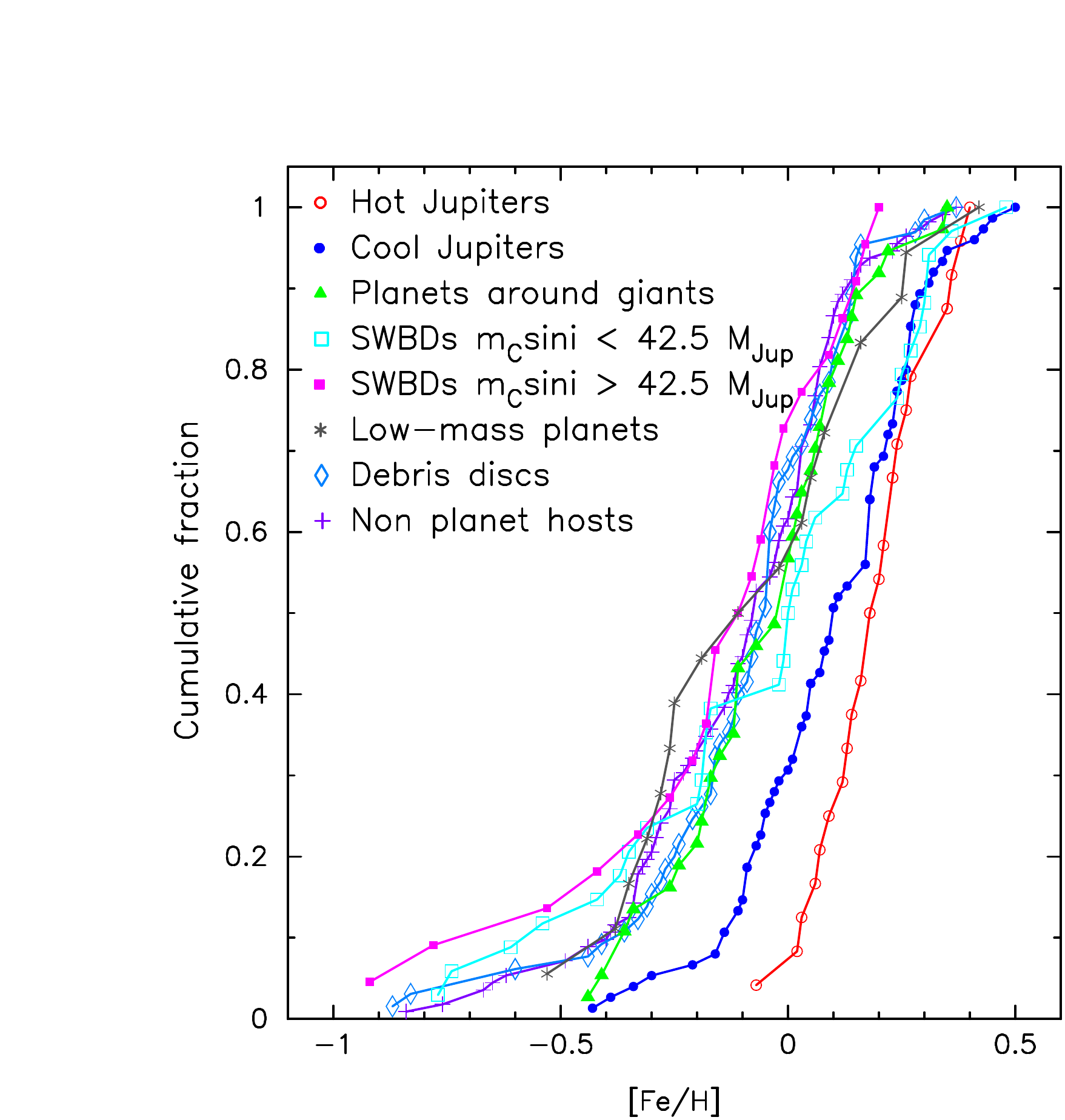}
\caption{
[Fe/H] cumulative  frequencies for the different samples analysed in this work. 
}
\label{cumu_distributions}
\end{figure}

 Figure~\ref{cumu_distributions} compares the cumulative distribution function
 of the stellar metallicity of the different stars analysed in this work.
 The stars have been divided into stars hosting hot Jupiters
 (if planets are located at distances smaller than 0.1 au), stars hosting cool 
 distant-gas-giant planets, giant stars with planets,
 stars harbouring low-mass brown dwarfs (m$_{\rm C}\sin i$ < 42.5 M$_{\rm Jup}$),
 stars with high-mass brown dwarfs (m$_{\rm C}\sin i$ > 42.5 M$_{\rm Jup}$),
  stars harbouring only low-mass planets (m$_{\rm C}\sin i$ < 30 M$_{\oplus}$),
  and stars hosting only debris discs.
 The figure shows
 that while the metallicity distribution of stars with hot and cool gas-giant
 planets are shifted towards high metallicity values, this is not the case
 for the other samples, which show metallicity distributions consistent
 with that of the comparison sample (i.e., stars without substellar companions).
 
\begin{figure*}[htb]
\centering
\begin{minipage}{\linewidth}
\includegraphics[scale=0.30]{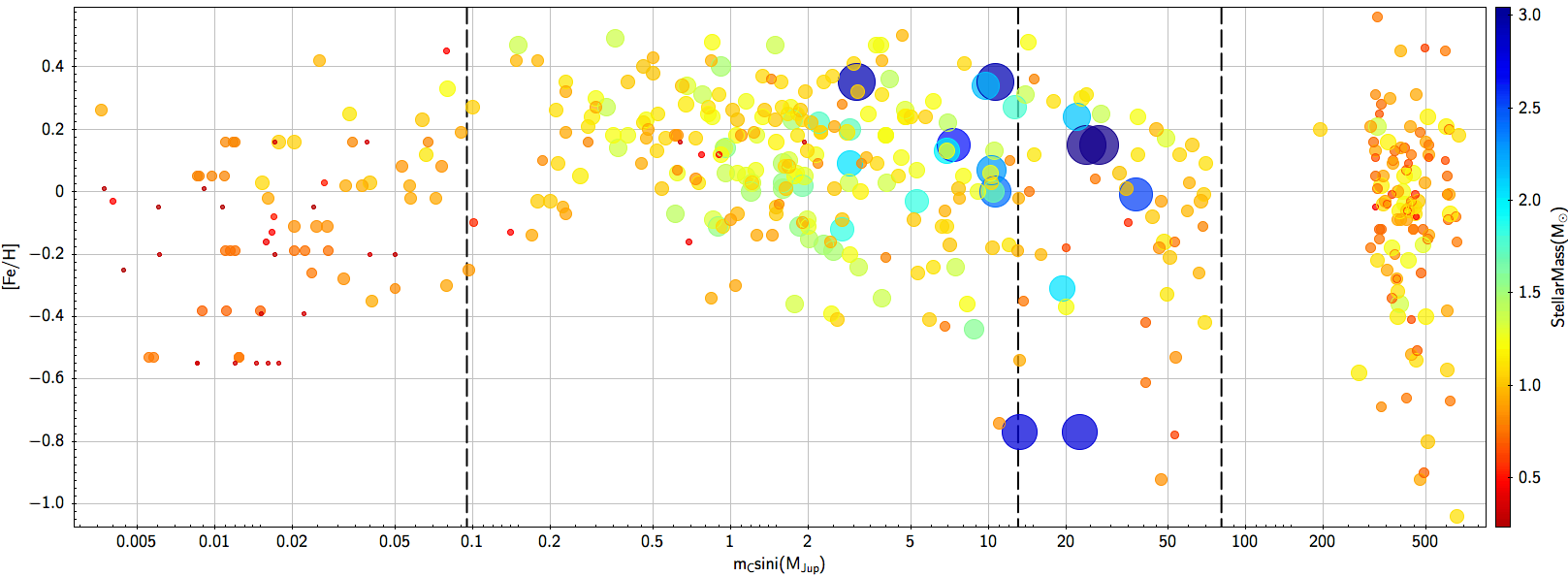} 
\end{minipage}
\caption{
Stellar metallicity of the host stars as a function of the minimum mass of the substellar companions.
Different colours and symbol sizes indicate the mass of the host star. Vertical dashed lines indicate
the standard mass loci of 
low-mass planets, gas-giant planets, brown dwarf, and stellar companions, from left to right
respectively.
}
\label{metal_mass}
\end{figure*}

 Figure~\ref{metal_mass} 
 shows the host star metallicity as a function of the (minimum)
 mass of the substellar companion. Colours and symbols indicate the mass of the host star.
 The figure clearly shows a tendency of lower host star's metallicities as the mass of the substellar
 companion increases. The figure also shows that more massive planets tend to orbit around more massive stars. 
 It is clear from the figure that 
 there is lack of planets around stars with metallicities lower
 than $\sim$ -0.4 dex. Stars with metallicities below this limit only harbour 
 substellar companions in the brown dwarf regime.
  We note that this result reefers to our sample and several planetary companions around more metal poor stars have
 been found.

 

 Our results suggest that there is a non universal planet formation
 mechanism. Different mechanisms may operate altogether and
 their relative efficiency change with the mass of the substellar object
 that is formed. For substellar objects with masses in the range
 30 M$_{\oplus}$ - 1 M$_{\rm Jup}$, high host star metallicities are found,
 suggesting that these planets are mainly formed by the core-accretion
 mechanism. As we move towards more massive substellar objects, the range
 of the host star metallicities increases towards more negative values, suggesting
 that a non-metallicity dependent formation mechanisms, such as gravitational instability
 or gravoturbulent fragmentation, might be at work.

 In a recent work, \cite{2018ApJ...853...37S} computed the mass at which substellar companions
 no longer preferentially orbit metal-rich stars finding that
 while objects with masses below 10 M$_{\rm Jup}$ orbit metal-rich
 stars, substellar companions with masses larger than 10 M$_{\rm Jup}$
 do not orbit metal-rich stars. We believe that our results are compatible with
 the findings by  \cite{2018ApJ...853...37S} showing that the most massive substellar
 objects tend to form like stars. 

\begin{figure}[htb]
\centering
\includegraphics[scale=0.22]{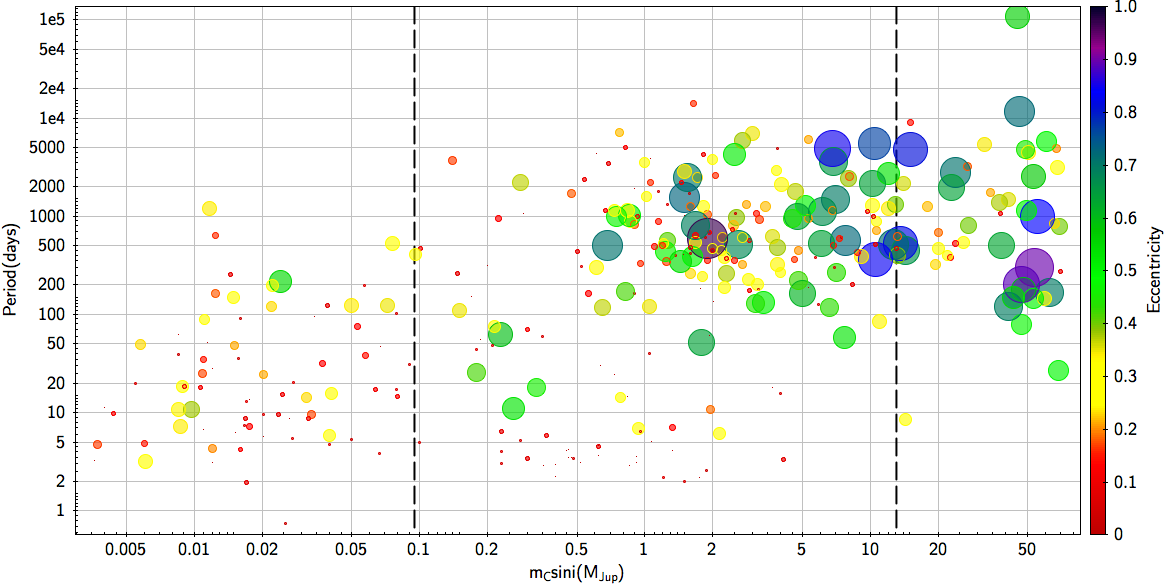}
\caption{ 
Orbital period as a function of the minimum mass. 
Different colours and symbol sizes indicate the eccentricity values. Vertical dashed lines indicate
the standard mass loci of 
low-mass planets, gas-giant planets, and brown dwarfs, from left to right. 
}
\label{period_eccentricity}
\end{figure}

 Figure~\ref{period_eccentricity} shows the orbital period of the substellar
 companions as a function of the (minimum) mass of the substellar companions. Different colours indicate
 the eccentricity.
 The figure shows that the more massive substellar companions show larger periods
 and eccentricities (P $>$ 100 days, $e$ $>$ 0.05). 
 On the other hand, less massive companions have shorter periods and a wider range of eccentricities.

 Setting together the trends from figures~\ref{metal_mass} and~\ref{period_eccentricity}
 it seems that as we move towards more massive planetary companions: 
 {\it i)} their host stars
 show a wider (towards negative values) range of metallicities and higher stellar masses;
 {\it ii)}  planets (or brown dwarfs) show longer periods and higher eccentricities.
 The differences in period and eccentricity distributions between both types of planets
 might be indicative of a different formation mechanism. 
 In addition, the trend with the host star metallicity
 suggests that the higher the mass of the substellar companion, the higher the
 probability that it is formed by a non-metallicity dependent formation mechanism.
 This general trend explain many of the correlations between the
 host star's metallicity and planetary properties discussed in recent works.

 \begin{itemize}

 \item {\it More massive stars host more massive planets.}
 It has been noticed that giant stars host more massive planets than
 their main-sequence counterparts \citep[e.g.][]{2007ApJ...665..785J,2007A&A...472..657L,2013A&A...554A..84M},
 although this result should be taken with caution as the detection of small planets
 around evolved stars is hampered by the large levels of stellar jitter in these
 stars \citep[e.g.][]{2016A&A...585A..73N}.
 We note that in our sample, stars in the mass range 1.5-2 M$_{\odot}$ host only planets
 with masses around 1 M$_{\rm Jup}$, while for stars more massive than 2 M$_{\odot}$ planets
 are more massive than 2 M$_{\rm Jup}$.
 This trend might reflect a correlation between disc gas masses and giant planet masses
 \citep{2011A&A...526A..63A,2012A&A...541A..97M} 
 as high-mass stars are likely to harbour more massive protoplanetary disk
 \citep[e.g.][]{2000prpl.conf..559N}. 
 In this scenario giant planet formation can occur in low metallicity but high-mass
 protoplanetary discs as it is the 
 amount of metals in the disc the factor that drives the planet formation process
 \citep[e.g.][]{2018ApJ...860..109G}. 
 The metallicity effect would depend on the mass of the disc, being the minimum metallicity
 required to for a massive planet lower for massive stars.
 \cite{2018ApJ...860..109G} found that  the relation between the amount of metals in the
 protoplanetary disc and the formation of giant planets does
 almost follow a linear relationship.
 
 The lack of a clear planet-metallicity correlation found for giant stars might be explained by the fact that
 they host more massive planets and these planets might find a way in their more massive planetary discs
 to bypass the core-acretion mechanism and form more like stars. 
 Finally, we note a tendency of more massive giants stars with substellar companions
 to have higher metallicities 
 in agreement with previous works \citep{2013A&A...554A..84M,2015A&A...574A..50J}.

 \item {\it Trends in brown dwarfs hosts.} 
 \cite{2014MNRAS.439.2781M,2014A&A...566A..83M}  showed that unlike gas-giant planet hosts,
 stars with brown dwarfs do not show metal-enrichment.
 \cite{2017A&A...602A..38M} found that stars with low-mass brown dwarfs
 tend to show higher metallicities
 than stars hosting more massive brown dwarfs.
 \cite{2014MNRAS.439.2781M,2017A&A...602A..38M} also discussed differences in the
 period-eccentricity distribution of massive and low-mass brown dwarfs.
 This result fits well with our
 interpretation that more massive substellar objects tend to form more like
  stars. 

 \item {\it Close-in and more distant planets.}
 Recent works \citep{2004MNRAS.354.1194S,2017arXiv171201035M,2018AJ....155...68W}
 have discussed whether hot Jupiters host stars show higher
 metallicities than more distant planets. As more distant planets are more
 massive than hot Jupiters  \citep{2007A&A...464..779R,2017A&A...604A..83B,2017A&A...603A..30S,2017MNRAS.466..443J,2017arXiv171201035M},
 see also Figure~\ref{period_eccentricity}, 
 they tend to orbit stars with a wider range of metallicities.

 \item {\it Planets around low-mass stars.}
 Planets around low-mass stars (M$_{\star}$ $<$ 1 M$_{\odot}$) are mainly
 low-mass planets and their host stars do not show metal enrichment. 
 They have short periods and low eccentricities. 
 On the other hand, very few gas-giant planets have been found orbiting around
 low-mass stars, showing their host stars metal-enrichment \citep{2013A&A...551A..36N}.
 We caution that these results refer to radial velocity planets. 
  Results from transit surveys are discussed in Sect.~\ref{kepler_comp}.

\end{itemize}

  We do not expect the general metallicity trends discussed in this work for massive planets and brown dwarfs
 to be severely affected by the different detection limits achieved for the different planet hosts.
 As discussed in Sect.~\ref{completness} planets of the mass of Jupiter at short periods can be detected in
 more than 95\% of our targets. More distant substellar companions  ($P$ $>$ 100 days) might be detected in a significant large percentage
 of our stars, between 60 and 80\%.
 The possible trend between  massive planetary companions and large periods might however be affected by our
 lower sensitivity to detect small gaseous planets (with masses of the order of the mass of Jupiter) at very long periods
 ($P$ $>$ 2000 days).  

 On the other hand, the results regarding low-mass planets should be regarded with caution as with the radial velocity data at hand,
 these planets can only be detected at short periods and around a small fraction of our stars (2 to 20\% according to our
 conservative simulations, see Figure~\ref{detectability}).

\section{The planet-metallicity correlation in context}\label{pm_context}

 In order to discuss our results into a broader context, 
 data from low-mass binaries have been included in Figure~\ref{metal_mass}. 
 The data is taken from \cite{2013AJ....145...52M}
 who compiled the metallicity of the primary stars mainly from high-resolution
 spectra. The mass of the late-K, M companions as well as the primaries are estimated by
 using a spectral type - stellar mass relationship based on the data
 provided by \cite{1977asqu.book.....A}.


 The figure shows that the tendency of a wider range of metallicities (lower values) 
 towards more massive objects continues in the low-mass stellar range. 
 Despite the fact we are comparing the minimum mass of substellar
 objects with estimates of the mass of stars, the trend that
 more massive substellar or stellar objects tend to form in a non-metallicity dependent 
 mechanism seems to hold, i.e., there seems to be a continuity between substellar
 and stellar companions. 
 According to this figure the core-accretion mechanism for planet formation
 would have its highest efficiency for forming planets with masses around 1 Jupiter's mass
 (hot Jupiters).

 It has been shown that the fraction of close 
 binaries of solar-type stars decreases with the
 metallicity while the 
 wide binary fraction is basically constant with metallicity at large separations
 \citep[e.g.][]{2018arXiv180802116M,2018arXiv180906860E}.
 Following the reasoning of \cite{2018arXiv180802116M} 
 massive and close substellar companions might form by fragmentation of the protostellar disc. 
 Protostellar discs of solar-type stars are usually optically thick and lower metallicities
 imply lower opacities and enhanced cooling rates which translate in higher probabilities
 of disc fragmentation\footnote{Although for very low metallicities, the disc becomes optically thin
 and the effect of lower metallicity would be the opposite.}.
 On the other hand, massive and distant companions might form by turbulent fragmentation of molecular
 cores, a process which is known to be independent of metallicity. 





 Figure~\ref{metal_mass} also reveals a 
 possible tendency of wider metallicities
 towards low-mass planets
 which may still be formed by core-accretion
 around low-metallicity stars. The low-metallicity environment implies long times for
 forming a core able to accret gas before the disc's dissipation, so only small planets
 and planetesimals can be formed \citep[e.g.][]{2012A&A...541A..97M}.
 However, the sample of M dwarfs planet hosts is still too small to make a strong
 claim in these sense.
  We also should note our limited sensitivity to low-mass planet's detection.
 Debris disc's masses do not help either as they are usually
 unavailable and subject to many assumptions. 

      



\section{Comparison with Kepler results}\label{kepler_comp}

 Given that our planet host sample is mainly selected from radial velocity surveys
 a comparison with the results from the {\sc Kepler} mission is mandatory in order to
 achieve a full vision of planet formation. 

 In a recent work, \cite{2018AJ....155...89P} analyse a large sample of {\sc Kepler}
 objects of interest with metallicities derived from spectroscopic observations
 finding that planets smaller than Neptunes (R$_{\rm P}$ $<$  4 R$_{\oplus}$) 
 are found around stars with a wide range of metallicities. On the other hand,
 sub-Saturns and Jupiters are found around metal-rich stars (their Figure~3). The authors also
 note a gradual upward trend in mean host star metallicity from smaller to larger
 planets in agreement with previous analysis of smaller samples  
 \citep{2012Natur.486..375B,2014Natur.509..593B}.
 These results support our findings that only stars hosting Jupiter-like planets
 show preferentially the metal-rich signature. 
 As lower planetary radius implies lower planetary masses, 
 although the relationship is complex and depend on the planet composition \citep[e.g.][]{2014ApJ...792....1L},
 we conclude that the {\sc Kepler} data supports 
 our suggestion that a tendency of lower metallicities towards low-mass planets
 might be hidden in 
 Fig.~\ref{metal_mass} as discussed in Sect.~\ref{pm_context}.

 Similar results have been found by \cite{2018arXiv180908385N} 
 who show that the host star metallicity, increases with
 larger planetary radius/mass up to about 1 M$_{\rm Jup}$ or 4 R$_{\oplus}$.
 For planetary masses larger than 4 M$_{\rm Jup}$ the authors also found
 that more massive planets have on average lower host star metallicities
 in agreement with our findings. 
 The authors also discuss that hot transiting planets
 (periods less than 10 days) orbit around stars with higher average
 metallicity in agreement with our previous results \citep{2017arXiv171201035M}
 and this work. 


 Studies of the {\sc Kepler} occurrence rates \citep{2015ApJ...798..112M,2015ApJ...814..130M}
 have confirmed that small planets (1.0 - 3.0 R$_{\oplus}$)  are more common around M dwarfs
 than around main-sequence FGK stars \citep{2012ApJS..201...15H}. At larger planetary
 radii planets become more common around sun-like stars. 
 Despite begin different samples (1.0 - 3.0 R$_{\oplus}$ planets
  correspond to masses below $\sim$ 8 M$_{\oplus}$, i.e., planets smaller than Neptune),
 a similar tendency of a larger occurrence of small planets towards less massive stars is found in our results from Fig.~\ref{metal_mass} where
 it can be seen that the vast majority of the low-mass planets 
 (m$_{\rm C}$$\sin i$ $<$ 30M$_{\oplus}$) orbit around stars with masses
 below 1 $M_{\odot}$. 




\section{Conclusions}\label{conclusions}

 Achieving a full vision of how planets and planetary systems form and evolve is only possible
 by analysing in a homogeneous way large samples of stars covering the full domain
 of parameters, i.e, including the different outcomes of the planet formation process
 (from planetesimals to massive brown dwarfs and low-mass stars) as well
 as the full range of host star's masses and types.
 In this work we performed a detailed analysis of the planet-metallicity correlation
 by analysing in a joined way the data from our previous works, focused on certain
 types of stars and/or planets. Most of the studied stars (excluding the M dwarf subsample)
 was analysed in the same way using similar spectra and techniques.

 Our results show a continuity between the formation of substellar and stellar companions
 driven by the metallicity of the host star. 
 The core-accretion formation mechanism would achieve its maximum efficiency
 for planets with masses between $\sim$ 0.2 and 2 M$_{\rm Jup}$. For more massive substellar
 objects as well in low-mass binary companions
 the range of host star's metallicities increases towards lower values, suggesting that
 both kind of objects tend to share similar formation mechanisms. 


 Another tendency towards lower host star's metallicities seems to be present towards
 the less massive outcomes of the planet formation process (low-mass planets and probably planetesimals)
 which may still be formed by the core-accretion method. However, this tendency might need
 additional confirmation. 
 

\begin{acknowledgements}

  This research was supported by the Italian Ministry of Education,
  University, and Research  through the
  \emph{PREMIALE WOW 2013} research project under grant
  \emph{Ricerca di pianeti intorno a stelle di piccola massa}.
  J. M. acknowledges support from the Ariel ASI-INAF agreement N. 2015-038-R.0.
  E. V., and C. E. acknowledge support from the \emph{On the rocks} project
  funded by the Spanish Ministerio de Econom\'ia y Competitividad
  under grant \emph{AYA2014-55840-P}.
  We sincerely appreciate the careful reading of the manuscript and the constructive comments of an anonymous referee.

\end{acknowledgements}

%
\bibliographystyle{aa}
\bibliography{metal_planet_full.bib}


\begin{appendix}
\section{Additional tables}



\longtab[1]{
\begin{longtable}{llcccccl} 
\caption{
 {\bf Basic properties of the stars considered in this work.}
  }\label{parameters_table_full}\\
\hline
 HIP/Other & HD &  V    &  SpType &   [Fe/H] & M$_{\star}$   & Ref$^{\dag}$ & Sample/Notes \\
     &    & (mag) &         &   (dex)  &  (M$_{\odot}$) &              &              \\
\hline
\endfirsthead
\caption{Continued.} \\
\hline
 HIP/Other & HD &  V    &  SpType &   [Fe/H] & M$_{\star}$   & Ref$^{\dag}$ & Sample/Notes \\
      &    & (mag) &         &   (dex)  &  (M$_{\odot}$) &              &              \\
\hline
\endhead
\hline
\endfoot
\hline
\endlastfoot
\hline\noalign{\smallskip}
171	&	224930	&	5.80	&	G3V	&	-0.83	$\pm$	0.01	&	0.68	$\pm$	0.00	&	1	&	Debris disc	\\
439	&	225213	&	8.60	&	M2.5	&	-0.27	$\pm$	0.09	&	0.33	$\pm$	0.07	&	5	&	M dwarf, comparison	\\
490	&	105	&	7.51	&	G0V	&	-0.16	$\pm$	0.03	&	0.99	$\pm$	0.02	&	1	&	Debris disc	\\
522	&	142	&	5.70	&	G1IV	&	0.07	$\pm$	0.02	&	1.25	$\pm$	0.01	&	4	&	Debris disc and two planets	\\
544	&	166	&	6.07	&	K0V	&	0.15	$\pm$	0.02	&	1.00	$\pm$	0.01	&	1	&	Debris disc	\\
729	&	448	&	5.57	&	G9III	&	0.04	$\pm$	0.04	&	1.81	$\pm$	0.08	&	2	&	Giant star, comparison sample	\\
873	&	645	&	5.84	&	K0III	&	0.03	$\pm$	0.02	&	1.78	$\pm$	0.05	&	2	&	Giant star, comparison sample	\\
910	&	693	&	4.89	&	F5V	&	-0.40	$\pm$	0.01	&	1.02	$\pm$	0.01	&	1	&	Comparison sample	\\
\end{longtable}
\tablefoot{
 $^{\dag}$ Reference for metallicity and stellar mass:
  {\bf(1)} \cite{2015A&A...579A..20M}; {\bf(2)} \cite{2016A&A...588A..98M}; {\bf(3)} \cite{2017A&A...602A..38M}; 
 {\bf(4)} \cite{2017arXiv171201035M}; {\bf(5)} \cite{2015A&A...577A.132M};  $^{\ddag}$  Value from the NASA exoplanet archive.
 Specifically from the summary of stellar information table for all stars, but KOI 415 for which we took the value from the KOI stellar properites table;
$^{\rm a)}$ \cite{2007A&A...472..657L}; $^{\rm b)}$ \cite{2016A&A...585A..46B};
$^{\rm c)}$ \cite{2009ApJ...707..768N}. Mass uncertainties listed as ``0.00'' should be understood as lower than 0.01 solar masses. \\
}%
}


\end{appendix}

\end{document}